\documentclass[showpacs,showkeys]{revtex4}
\usepackage{graphicx}
\usepackage{epstopdf}

\usepackage{amsmath}
\usepackage{amssymb}
\usepackage{color}

\begin{document}

\title{Asymptotically flat gravitating spinor field solutions. Step 3 - Einstein-Dirac equations: cylindric case}

\author{Vladimir Dzhunushaliev}
\email{vdzhunus@krsu.edu.kg}
\affiliation{Institute for Basic Research,
Eurasian National University,
Astana, 010008, Kazakhstan \\
and \\
Institute of Physics of National Academy of Science
Kyrgyz Republic, 265 a, Chui Street, Bishkek, 720071,  Kyrgyz Republic}

\date{\today}

\begin{abstract}
Einstein-Dirac equations for two spinor fields are considered. It is shown that in this case one can obtain self-consistent equations set for these gravitating spinors. The key idea to obtain Einstein-Dirac equations is to use special ans\"atze for spinor fields in order to zero some off-diagonal componentes of the spinor energy - momentum tensor.
\end{abstract}

\pacs{04.40.-b}
\keywords{Einstein-Dirac equations}

\maketitle

\section{Introduction}

Self-consistent solutions of Einstein-Dirac equations are a \textcolor{blue}{\emph{terra incognita}} in general relativity.  The problem in the construction of such solutions in general relativity with a spinor field is that the spinor field has a spin that leads to off-diagonal components of the energy-momentum tensor and consequently to an off-diagonal metric. Note that the solutions for a spinor field propagating on a curved background are known (for details, see Ref. \cite{chandrasekhar}). The cosmological solutions with a spinor field are known as well, see Refs. \cite{Saha:2010zza}-\cite{Ribas:2010zj}.

Here we will consider cylindrically symmetric solution in Einstein-Dirac gravity with two noninteracting Weyl spinor fields. We use two spinor fields because in this case one can choose two spinor ans\"atze in such a way that the off-diagonal components of energy-momentum tensor from every spinor field have the opposite sign and consequently will annihilate each other.

\section{Einstein-Dirac equations}

Our main goal is to find cylindrically symmetric solution for Einstein-Dirac equations
\begin{eqnarray}
    R_{\mu \nu} - \frac{1}{2} g_{\mu \nu} R &=& \textcolor{red}{-} \varkappa T_{\mu \nu},
\label{2-10}\\
    \sigma^\mu_{(a)(b')} P^{(a)} + i m \bar{Q}^{(c')} \epsilon_{(c')(b')} &=& 0,
\label{2-20}\\
    \sigma^\mu_{(a)(b')} Q^{(a)} + i m \bar{P}^{(c')} \epsilon_{(c')(b')} &=& 0
\label{2-30}
\end{eqnarray}
where $R_{\mu \nu}$ is the Ricci tensor; $R$ is the scalar curvature;
$g_{\mu \nu}$ is the metric; $\mu, \nu = 0,1,2,3$;
$T_{\mu \nu} = (T_P)_{\mu \nu} + (T_Q)_{\mu \nu}$, $(T_{P,Q})_{\mu \nu}$ are the energy-momentum for the Weyl spinors $P$ and $Q$; $\sigma^\mu_{(a)(b')}$ are the Pauli matrixes in a dyad formalism in a curve space;
$(a), (b') = (0), (1)$ are the dyad indices; $P^{(a)}, Q^{(a)}$ are two spinors in a dyad representation; $\bar P^{(a)}, \bar Q^{(a)}$ are complex conjugated spinors; the Levi-Civita symbols $\epsilon_{(c)(b)}, \epsilon_{(c')(b')}$ and $\epsilon^{(c)(b)}, \epsilon^{(c')(b')}$ are the metrics in the spinor space for lowering and raising the spinor indices; $m$ is the mass. The sign $\textcolor{red}{(-)}$ in the front of RHS of Eq. \eqref{2-20} is chosen as in Ref. \cite{chandrasekhar} but it is necessary to check this sign. 

We will use Newman-Penrose formalism for the spinor analysis (for details, see Appendix \ref{np}).

We will consider the simplest case $m=0$ that leads to: (a) equations for every spinor field $P$ and $Q$ will be separated; (b) one can choose ans\"atze for both spinor fields in such a way that \textcolor{blue}{\emph{some off-diagonal components of the energy-momentum tensor will be zero}}; (c) equations for every spinor field $P$ and $Q$ will be \textcolor{blue}{\emph{the same}} and (d) consequently the spinors $P$ and $Q$  will be \textcolor{blue}{\emph{the same}}.

We seek the solution in the following form: the metric
\begin{equation}
    ds^2 = e^{2 \nu(\rho) - 2 \mu(\rho)} \left(
    	dt^2 - d \rho^2
    \right) - w^2(\rho) e^{-2 \mu(\rho)} d \varphi^2 -
    e^{2 \mu(\rho)} \left[
    	dz + A(\rho) d \varphi
    \right]^2,
\label{2-40}
\end{equation}
the spinors
\begin{eqnarray}
  P^{(a)} &=& e^{i \omega t} \left(
	\begin{array}{c}
		p_1 \\
		p_2
	\end{array}
	\right),
\label{2-50}\\
  Q^{(a)} &=& e^{i \omega t} \left(
	\begin{array}{c}
		q_2 \\
		q_2
	\end{array}
	\right).
\label{2-60}
\end{eqnarray}
Later we will see that a specific choice of $q_{1,2}$ allows us to kill some off-diagonal components of the spinor energy-momentum.

The null vectors that are necessary to the calculations of $R_{\mu \nu}, R, \sigma^\mu_{(a)(b')}$ and so on are (for details, see Appendix \ref{np}) 
\begin{eqnarray}
  l^{\mu} &=& \frac{1}{\sqrt{2}} \left(
	   e^{\mu -\nu}, 0,0, e^{-\mu}
	\right),
\label{2-62}\\
  n^{\mu} &=& \frac{1}{\sqrt{2}} \left(
	   e^{\mu -\nu}, 0,0, -e^{-\mu}
	\right),
\label{2-64}\\
  m^{\mu} &=& \frac{1}{\sqrt{2}} \left(
	   0, -i e^{\mu - \nu}, -\frac{e^{\mu}}{w}, \frac{A e^{\mu}}{w}
	\right),
\label{2-66}\\
  \bar m^{\mu} &=& \frac{1}{\sqrt{2}} \left(
	   0, i e^{\mu - \nu}, -\frac{e^{\mu}}{w}, \frac{A e^{\mu}}{w}
	\right).
\label{2-68}
\end{eqnarray}
The LHS's of Einstein equations \eqref{2-10} are
\begin{eqnarray}
    G_{00} &=& - \frac{w''}{w} + \frac{w' \nu'}{w} - {\mu'}^2 -
    e^{4 \mu} \frac{{A'}^2}{4 w^2},
\label{2-70}\\
    G_{11} &=& \frac{w' \nu '}{w} - {\mu '}^2 - e^{4 \mu} \frac{{A'}^2}{4 w^2},
\label{2-90}\\
    G_{22} &=& e^{-2 \nu} \left(
    	w^2 + e^{4 \mu} A^2
    \right) \nu '' -
    2 e^{4 \mu - 2 \nu} A^2 \mu '' -
    e^{4 \mu - 2 \nu} A A'' +
    e^{4 \mu - 2 \nu} A^2 \frac{w''}{w} +
\nonumber \\
    &&
    e^{-2 \nu} \left(
    	w^2 + e^{4 \mu} A^2
    \right) {\mu'}^2 +
    e^{4 \mu - 2 \nu} A^2 \left(
    	\frac{w' A'}{w A} - \frac{1}{4} \frac{{A'}^2}{A^2} +
    	\frac{3}{4} e^{4 \mu} \frac{{A'}^2}{w^2} - 2 \frac{w' \mu '}{w} -
    	4 \frac{A' \mu '}{A}
    \right) ,
\label{2-100}\\
    G_{23} &=& G_{32} = e^{4 \mu - 2 \nu} A \left(
    	\nu '' - 2 \mu '' - \frac{A''}{2 A} + \frac{w''}{w}
    \right) + A e^{4 \mu - 2 \nu} \left(
    	\frac{1}{2} \frac{A' w'}{A w} +
    	\frac{3}{4} e^{4 \mu} \frac{{A'}^2}{w^2} -
    	2 \frac{w'\mu '}{w} - 2 \frac{A' \mu'}{A} + {\mu'}^2
    \right),
\label{2-110}\\
    G_{33} &=& e^{4 \mu - 2 \nu} \left(
    	\nu '' - 2 \mu'' + \frac{w'}{w}
    \right) + e^{4 \mu - 2 \nu} \left(
    	\frac{3}{4} e^{4 \mu} \frac{{A'}^2}{w^2} -
    	2 \frac{w' \mu'}{w} + {\mu'}^2
    \right)
\label{2-120}
\end{eqnarray}
where $d(\cdots)/d\rho = (\cdots)^\prime$. The energy-momentum tensor can be found using \eqref{a1-107} and \eqref{a1-220} and it has following nonzero off-diagonal components of the energy-momentum for the $P$ spinor: $(T_P)_{02}, (T_P)_{03}, (T_P)_{23} \neq 0$. The same for the $Q$ spinor: $(T_Q)_{02}, (T_Q)_{03}, (T_Q)_{23} \neq 0$. The key idea is that if we choose $q_1 = -p_2, q_2 = p_1$ then
\begin{equation}
  (T_P)_{02} = - (T_Q)_{02} , \quad
  (T_P)_{03} = - (T_Q)_{03}
\label{2-130}
\end{equation}
and nonzero components are following
\begin{equation}
  (T_P)_{23} , (T_Q)_{23} \neq 0 .
\label{2-140}
\end{equation}
Consequently we have following energy-momentum for two spinor fields
\begin{eqnarray}
    T_{00} &=& e^{\nu - \mu} \omega \left( p_1^2 + p_2^2 \right),
\label{2-150}\\
    T_{11} &=& e^{\nu - \mu} \left( p^\prime_1 p_2 - p_1 p^\prime_2 \right),
\label{2-160}\\
    T_{22} &=& - A^2 \frac{e^{\mu - \nu}}{2 \sqrt{2}} \left( p_1^2 + p_2^2 \right) \left[
        2 \frac{w'}{A} + e^{4 \mu} \frac{A'}{w} - \frac{w A'}{A^2} -
        4 \frac{w \mu'}{A}
    \right],
\label{2-170}\\
    T_{23} &=& T_{32} = - A \frac{e^{\mu - \nu}}{2 \sqrt{2}} \left( p_1^2 + p_2^2 \right)
    \left(
    	\frac{w'}{A} + e^{4 \mu} \frac{A'}{w} - 2 \frac{w \mu'}{A}
    \right),
\label{2-180}\\
    T_{33} &=& e^{5 \mu - \nu} \frac{A'}{w} \left( p_1^2 + p_2^2 \right).
\label{2-190}
\end{eqnarray}
Finally Dirac equations \eqref{2-20} \eqref{2-30} have the form
\begin{eqnarray}
    p^\prime_1 + \frac{p_1}{2} \left(
        \frac{w'}{w} - \mu' + \nu^\prime
    \right) + p_2 \left(
        - \omega + e^{2 \mu} \frac{A'}{4w}
    \right)&=& 0,
\label{2-200}\\
    p^\prime_2 + \frac{p_2}{2} \left(
        \frac{w'}{w} - \mu' + \nu^\prime
    \right) + p_1 \left(
        \omega - e^{2 \mu} \frac{A'}{4w}
    \right)&=& 0.
\label{2-210}
\end{eqnarray}

\section*{Acknowledgements}

I am grateful to the Research Group Linkage Programme of the Alexander von Humboldt Foundation for the support of this research.

\appendix

\section {Newman-Penrose formalism}
\label{np}

The Newman-Penrose formalism can be found in Ref's \cite{chandrasekhar} \cite{frolov}. But there are many misprints and confusions between dyad $(a)$ and spinor $A$ indices. In this Appendix we follow to Ref's \cite{chandrasekhar} \cite{frolov} and will try to avoid misprints and confusions.

Weyl spinors $\xi^A$ and $\eta^{A'}$ of rank 1 are complex vectors in a 2D space ($A, A' = 0,1$) subject to the transformations
\begin{eqnarray}
    \xi_*^A &=& \alpha^A_{\phantom{A}B} \xi^B ,
\label{a1-10}\\
    \eta_*^{A'} &=& \bar \alpha^{A'}_{\phantom{A'}B'} \xi^{B'}
\label{a1-20}
\end{eqnarray}
where $A,B,A',B' = 0,1$ and $\alpha^A_{\phantom{A}B}, \bar \alpha^{A'}_{\phantom{A'}B'}$ are complex conjugate matrices with unit determinants
\begin{equation}
    \det \alpha^A_{\phantom{A}B} = \det \bar \alpha^{A'}_{\phantom{A'}B'} = 1.
\label{a1-30}
\end{equation}
The metrics $\epsilon_{AB}, \epsilon_{A' B'}$ and $\epsilon^{AB}, \epsilon^{A' B'}$ are used for lowering and raising spinor indices
\begin{eqnarray}
    \xi_A &=& \xi^C \epsilon_{CA}, \quad \xi_{A'} = \xi^{C'} \epsilon_{C'A'},
\label{a1-40}\\
    \xi^A &=& \epsilon^{AC} \xi_C, \quad \xi^{A'} = \epsilon^{A'C'} \xi_{C'}
\label{a1-50}
\end{eqnarray}
At each point of the spacetime one can set up an orthonormal dyad basis $\zeta^A_{(a)}, \zeta^{A'}_{(a')}$ with $(a), (a') = (0), (1), A, A' = 0,1$ for spinors. One can raise and lower dyad indices by
$\epsilon^{(a)(b)}, \epsilon^{(a')(b')}$ and $\epsilon_{(a)(b)}, \epsilon_{(a')(b')}$
\begin{eqnarray}
    \xi_{(a)} &=& \xi^{(c)} \epsilon_{(c)(a)}, \quad \xi_{(a')} = \xi^{(c')} \epsilon_{(c')(a')},
\label{a1-60}\\
    \xi^{(a)} &=& \epsilon^{(a)(c)} \xi_{(c)}, \quad \xi^{(a')} = \epsilon^{(a')(c')} \xi_{(c')}.
\label{a1-70}
\end{eqnarray}
We can project any spinor $\xi_A$ on the dyad basis
\begin{equation}
    \xi_{(a)} = \xi_A \zeta^A_{(a)}, \quad \xi_{(a')} = \xi_{A'} \zeta^{A'}_{(a')}
\label{a1-80}
\end{equation}
If we introduce null vectors (isotropic tetrad) $l^\mu, n^\mu, m^\mu, \bar m^\mu$ then we can define the Pauli matrixes in the dyad representation in a curve spacetime
\begin{eqnarray}
    \sigma^\mu_{(a)(b')} &=& \frac{1}{\sqrt{2}}
    \left(
     \begin{array}{cc}
     l^\mu         & m^\mu \\
     \bar m^\mu    & n^\mu
     \end{array}
    \right),
\label{a1-90}\\
    \sigma_\mu^{(a)(b')} &=& \frac{1}{\sqrt{2}}
    \left(
     \begin{array}{cc}
     n_\mu     & -\bar m_\mu \\
     -m_\mu    & l_\mu
     \end{array}
    \right).
\label{a1-100}
\end{eqnarray}
The Pauli matrices gives us the one-to-one correspondence between 4-vectors $X^\mu$ and spinor of rank two $X^{(a)(b')}$
\begin{eqnarray}
    X^\mu &=& \sigma^\mu_{(a)(b')} X^{(a)(b')} , \quad
    X^{(a)(b')} = \sigma_\mu^{(a)(b')} X^\mu ,
\label{a1-105}\\
    X_\mu &=& \sigma_\mu^{(a)(b')} X_{(a)(b')} , \quad
    X_{(a)(b')} = \sigma^\mu_{(a)(b')} X_\mu .
\label{a1-107}
\end{eqnarray}
The covariant derivative of a spinor field $\xi_a$ is defined as
\begin{equation}
    \nabla_\mu \xi_{(a)} = \partial_\mu \xi_{(a)} + \Gamma^{(b)}_{\mu (a)} \xi_{(b)}
\label{a1-110}
\end{equation}
where
\begin{eqnarray}
    \Gamma^{(a)}_{\mu (b)} &=& \frac{1}{2} \sigma_\lambda^{(a)(b')}
    \left[
        \sigma^\nu_{(b)(b')} \Gamma^\lambda_{\mu \nu} +
        \partial_\mu \sigma^\lambda_{(b)(b')}
    \right],
\label{a1-120}\\
    \bar{\Gamma}^{(a')}_{\mu (b')} &=& \overline{\Gamma^{(a)}_{\mu (b)}}
\label{a1-130}
\end{eqnarray}
where $\overline{(\cdots)}$ is the complex conjugation. According to \eqref{a1-107} one can introduce the covariant derivative $\nabla_{(a)(b')}$
\begin{equation}
    \nabla_{(a)(b')} = \sigma^\mu_{(a)(b')} \nabla_\mu .
\label{a1-140}
\end{equation}
One can introduce the spin coefficients with spinor indices only
\begin{eqnarray}
    \Gamma_{(a)(b)(c)(d')} &=& \sigma^\mu_{(c)(d')} \epsilon_{(f)(b)}
    \Gamma^{(f)}_{\mu (a)},
\label{a1-150}\\
    \bar \Gamma_{(a')(b')(c')(d)} &=& \sigma^\mu_{(d)(c')} \epsilon_{(f')(b')}
    \bar \Gamma^{(f')}_{\mu (a')}.
\label{a1-160}
\end{eqnarray}
The spin coefficients have following symmetry properties
\begin{equation}
    \Gamma_{(a)(b)(c)(d')} = \Gamma_{(b)(a)(c)(d')}, \quad
    \bar \Gamma_{(b')(a')(c')(d)}  = \bar \Gamma_{(a')(b')(c')(d)} .
\label{a1-170}
\end{equation}
The spin coefficients can be calculated as follows
\begin{equation}
    \Gamma_{(a)(b)(c)(d')} = \frac{1}{2} \epsilon^{(p')(q')}
     \sigma^\mu_{(a)(q')}  \sigma^\nu_{(c)(d')} \nabla_\nu
     \sigma_{\mu (b)(p')} .
\label{a1-175}
\end{equation}
In the Newman-Penrose formalism, these coefficients are assigned special symbols which are listed in the Table \ref{tb}. These definitions of the spin coefficients are in agreement with those defined in terms of the Ricci-rotations coefficients $\gamma_{\bar a \bar b \bar c}$.
\begin{table}[t]
\begin{center}
\begin{tabular}{|c|c|c|c|}
  \hline
  (c)(d$'$) & (a)(b)=(0)(0) & (0)(1),(1)(0) & 11 				    \\ \hline
  (0)(0$'$) & $\kappa$ 			& $\varepsilon$     & $\pi$ 	\\ \hline
  (1)(0$'$) & $\rho$ 				& $\alpha$ 			& $\lambda$ \\ \hline
  (0)(1$'$) & $\sigma$ 			& $\beta$ 			& $\mu$ 	\\ \hline
  (1)(1$'$) & $\tau$ 				& $\gamma$ 			& $\nu$ 	\\ \hline
\end{tabular}
\end{center}
\caption{The spin coefficients in Newman-Penrose formalism.
}
\label{tb}
\end{table}

In order to define the Ricci-rotations coefficients $\gamma_{\mu\nu\rho}$ we introduce the tetrad
$e^{\phantom{a} \mu}_{\bar a}$ in following way
\begin{equation}
    e^{\phantom{a} \mu}_{\bar 0} = l^\mu, e^{\phantom{a} \mu}_{\bar 1} = n^\mu,
    e^{\phantom{a} \mu}_{\bar 2} = m^\mu, e^{\phantom{a} \mu}_{\bar 3} = \bar m^\mu
\label{a1-180}
\end{equation}
where $\bar a = \bar 1, \bar 2, \bar 3, \bar 4$ are the tetrad indices. Minkowski metric for Newman-Penrose formalism is chosen as
\begin{equation}
   \eta_{\mu \nu} =
   \left(
     \begin{array}{cccc}
       0 & 1 & 0 & 0 \\
       1 & 0 & 0 & 0 \\
       0 & 0 & 0 & -1 \\
       0 & 0 & -1 & 0 \\
     \end{array}
   \right).
\label{a1-190}
\end{equation}
For the calculation of the Ricci-rotation coefficients one can introduce $\lambda_{\bar a \bar b \bar c}$
\begin{equation}
   \lambda_{\bar a \bar b \bar c} = \left(
    e_{\bar b \mu , \nu} - e_{\bar b \nu , \mu}
   \right) e^{\phantom{a} \mu}_{\bar a} e^{\phantom{a} \nu}_{\bar c}
\label{a1-200}
\end{equation}
and finally
\begin{equation}
   \gamma_{\bar a \bar b \bar c} = \frac{1}{2} \left(
    \lambda_{\bar a \bar b \bar c} - \lambda_{\bar b \bar c \bar a} +
    \lambda_{\bar c \bar a \bar b}.
   \right)
\label{a1-210}
\end{equation}
The spin coefficients are
\begin{eqnarray}
    \kappa &=& \gamma_{\bar 3 \bar 1 \bar 1}, \quad
    \rho = \gamma_{\bar 3 \bar 1 \bar 4}, \quad
    \varepsilon = \frac{1}{2} \left(
        \gamma_{\bar 2 \bar 1 \bar 1} + \gamma_{\bar 3 \bar 4 \bar 1}
    \right),
\label{a1-220}\\
    \sigma &=& \gamma_{\bar 3 \bar 1 \bar 3}, \quad
    \mu = \gamma_{\bar 2 \bar 4 \bar 3}, \quad
    \gamma = \frac{1}{2} \left(
        \gamma_{\bar 2 \bar 1 \bar 2} + \gamma_{\bar 3 \bar 4 \bar 2}
    \right),
\label{a1-230}\\
    \lambda &=& \gamma_{\bar 2 \bar 4 \bar 4}, \quad
    \tau = \gamma_{\bar 3 \bar 1 \bar 2}, \quad
    \alpha = \frac{1}{2} \left(
        \gamma_{\bar 2 \bar 1 \bar 4} + \gamma_{\bar 3 \bar 4 \bar 4}
    \right),
\label{a1-240}\\
    \nu &=& \gamma_{\bar 2 \bar 4 \bar 2}, \quad
    \pi = \gamma_{\bar 2 \bar 4 \bar 1}, \quad
    \beta = \frac{1}{2} \left(
        \gamma_{\bar 2 \bar 1 \bar 3} + \gamma_{\bar 3 \bar 4 \bar 3}
    \right).
\label{a1-250}
\end{eqnarray}
The Riman tensor is
\begin{equation}
   R_{\bar a \bar b \bar c \bar d} = - \gamma_{\bar a \bar b \bar c, \bar d} +
   \gamma_{\bar a \bar b \bar d, \bar c} + \gamma_{\bar b \bar a \bar f}
   \left(
   	\gamma_{\bar c \phantom{f} \bar d}^
   	{\phantom{\bar c} \bar f \phantom{\bar d}} -
   	\gamma_{\bar d \phantom{f} \bar c}^
   	{\phantom{\bar d} \bar f \phantom{\bar c}}
   \right) +
   \gamma_{\bar f \bar a \bar c}
   \gamma_{\bar b \phantom{f} \bar d}^
   	{\phantom{\bar b} \bar f \phantom{\bar d}} -
   	\gamma_{\bar f \bar a \bar d}
   \gamma_{\bar b \phantom{f} \bar c}^
   	{\phantom{\bar b} \bar f \phantom{\bar c}}
\label{a1-260}
\end{equation}
with usual definitions for the Ricci tensor $R_{\bar a \bar b}$
\begin{equation}
   R_{\bar a \bar b} = R^{\bar c}_{\phantom{c}\bar a \bar c \bar b}
\label{a1-264}
\end{equation}
and scalar curvature
\begin{equation}
   R = \eta^{\bar a \bar b} R_{\bar a \bar b}.
\label{a1-267}
\end{equation}
The spinor energy-momentum for the spinor $P_{(a)}$ is given by
\begin{equation}
   T_{{a}{a'}{b}{b'}} = \frac{1}{2} \left[
    P_{(a)} \nabla_{(b)(b')} \bar P_{(a')} -
    \bar P_{(a')} \nabla_{(b)(b')} P_{(a)} +
    P_{(b)} \nabla_{(a)(a')} \bar P_{(b')} -
    \bar P_{(b')} \nabla_{(a)(a')} P_{(b)}
   \right].
\label{a1-270}
\end{equation}

\end{document}